\documentclass[graybox,vecphys,natbib]{svmult}

\usepackage{mathptmx}       % selects Times Roman as basic font
\usepackage{helvet}         % selects Helvetica as sans-serif font
\usepackage{courier}        % selects Courier as typewriter font
\usepackage{makeidx}         % allows index generation
\usepackage{graphicx}        % standard LaTeX graphics tool
\usepackage{multicol}        % used for the two-column index
\usepackage[bottom]{footmisc}% places footnotes at page bottom

\makeindex             % used for the subject index
\usepackage{amsmath,amssymb}
\usepackage{url}

\newcommand{\dz}{\ensuremath{\,\mathrm{d}z}}

\newcommand{\Msun}{\ensuremath{\text{M}_{\odot}}}
\newcommand{\msy}{{\ensuremath{\Msun\,\text{year}^{\mbox{-1}}}}}
\newcommand{\dMdtl}{\ensuremath{{\text{d}M}/{\text{d}t}}}
\newcommand{\Teff}{\ensuremath{T_\text{eff}}}
\newcommand{\Kelvin}{\ensuremath{\text{K}}}
\newcommand{\kms}{\ensuremath{\text{km}\,\text{s}^{-1}}}

\newcommand{\mn}{\ensuremath{{\mu\nu}}}
\newcommand{\deriv}[2]{\frac{{\text d}#1}{{\text d}#2}}
\newcommand{\dm}{\ensuremath{\,\text{d}m}}
\newcommand{\dnu}{\ensuremath{\,\text{d}\nu}}
\newcommand{\velocity}{\ensuremath{v}}

\newcommand{\zav}[1]{\left(#1\right)}

\newcommand{\Jiri}{Ji\v{r}\'{\i}}
\newcommand{\Kubat}{Kub\'at}
\newcommand{\Surlan}{\v{S}urlan}

\newcommand{\TLUSTY}{{\tt TLUSTY}}

\DeclareMathAlphabet{\mathsc}{OT1}{cmr}{m}{sc}
\def\testbx{bx}%
\DeclareRobustCommand{\ion}[2]{%
\relax\ifmmode
\ifx\testbx\f@series
{\mathbf{#1\,\mathsc{#2}}}\else
{\mathrm{#1\,\mathsc{#2}}}\fi
\else\textup{#1\,{\mdseries\textsc{#2}}}%
\fi}

\begin{document}

\title*{NLTE analysis of spectra: OBA stars}
\author{{\Jiri} {\Kubat} \and Brankica {\Surlan}}
\authorrunning{{\Jiri} {\Kubat} \and Brankica {\Surlan}}
\institute{{\Jiri} {\Kubat} \at Astronomick\'y \'ustav AV \v{C}R,
Fri\v{c}ova 298, 251 65 Ond\v{r}ejov, Czech Republic, \email{kubat@sunstel.asu.cas.cz}
\and
Brankica {\Surlan} \at Astronomick\'y \'ustav AV \v{C}R, Fri\v{c}ova
298, 251 65 Ond\v{r}ejov, Czech Republic \and Matemati\v{c}ki Institut
SANU, Kneza Mihaila 36, 11001 Beograd, Republic of Serbia,
\email{surlan@sunstel.asu.cas.cz}
}
\maketitle

\abstract{
Methods of calculation of NLTE model atmosphere are discussed.
The NLTE trace element procedure is compared with the full NLTE model
atmosphere calculation.
Differences between LTE and NLTE atmosphere modeling are evaluated.
The ways of model atom construction are discussed.
Finally, modelling of expanding atmospheres of hot stars with winds is
briefly reviewed.}

%%%%%%%%%%%%%%%%%%%%%%%%%%%%%%%%%%%%%%%%%%%%%%%%%%%%%%%%%%%%%%%%%%%%%%%%
\section{Introduction}

As hot stars we usually consider stars of spectral types A, B, and O.
These stars are hotter than about $8500\,\Kelvin$ (this temperature
value depends on who is classifying them), they are also more massive
and have several times larger radius than the Sun.
Their spectra may be characterized by fewer spectral lines than cool
stars, and by the absence of molecular lines, which is given by their
higher temperature.
Also thanks to temperature the major convection zone caused by hydrogen
ionization is absent in their atmospheres.
In addition, strong NLTE effects appear in their atmospheres, which
influence not only line formation, but the whole atmospheric structure.

However, they also differ depending on their spectral type.
The atmospheres of A-type stars are usually very quiet with almost a
total absence of atmospheric motions.
Consequently, specific phenomena like chemical stratification of their
atmospheres may develop.
This is also the reason for strong chemical peculiarities found in many
of them.
The atmospheres of B-type stars are less quiet, and macroscopic
atmospheric motions start to appear there.
These stars are very often rapidly rotating, which causes development of
circumstellar envelopes and emission lines in their spectra (the case of
Be stars).
On the other hand, many stars of this type are pulsating (e.g. the
$\beta$ Cep type stars), some even show non-radial pulsations.
The O-type stars are typified by strong stellar winds, which cause
specific emission lines (with P-Cygni type profiles) in their spectra.
For all these spectral types, NLTE effects are essential for both
atmospheric structure and line formation.

Consequently, the NLTE approach for analysis of their spectra is
inevitable.
There are two basic possibilities to calculate stellar atmosphere models
with the departures from LTE.
The first and consistent one is calculation of full NLTE model
atmospheres including all chemical elements. 
However, this option is still computationally prohibitive, so we usually
restrict the number of chemical elements and include only those, which
are important for the determination of the atmospheric structure.
The second option is to solve the NLTE line formation problem for a
particular atom or ion (called a trace element) for a given (i.e. fixed)
model atmosphere.
This model may be calculated assuming either LTE or NLTE.
The former option saves significant amount of time necessary for
calculations.
On the other hand, LTE model atmospheres are physically inconsistent
\citep{Mihalas:1978, Hubeny:Mihalas:2014}.

%%%%%%%%%%%%%%%%%%%%%%%%%%%%%%%%%%%%%%%%%%%%%%%%%%%%%%%%%%%%%%%%%%%%%%%%
\section{Trace elements}

Solution of a NLTE line formation problem for trace elements splits the
solution of the problem in two steps.

The first step is a calculation of the model atmosphere, which includes
only those elements and their transitions, which are important for
determination of the atmospheric structure, specifically continua of
abundant elements and strong atomic lines.
Many weaker lines in the optical region may be omitted in this step,
since their influence on the atmospheric structure is negligible.
This model atmosphere may assume either LTE or NLTE, however, a NLTE
model atmosphere should be in principle always preferred in this step.

Having a model atmosphere we solve a NLTE line formation problem, i.e.
simultaneous solution of the radiative transfer equation and the
equations of statistical equilibrium for a selected element (the {\em
trace element}).
Since the model atmosphere (i.e. the full structure) is not solved in this
step, the model atom may be very detailed and include many lines and
continua important for formation of lines of this elements.

Several conditions have to be fulfilled to allow us to consider a
particular element as a trace element.
First of all, its influence on the atmospheric structure has to be
negligible.
Also the opacity of the trace element has to be negligible compared to
the opacity of non-trace elements, or NLTE effects on opacities of this
element have negligible influence on the atmospheric structure.
Trace element also should not be a significant source of free electrons.
Simply stated, any detailed trace element calculation \emph{must not}
influence the other elements.
If this happens, we have to improve the background model.
An example of a trace element may be, e.g., argon in B-type stars
\citep{Lanz:etal:2008}.
These authors used fully blanketed NLTE model atmospheres as background
models, which is the best available option.
More discussion about NLTE for trace elements can be found in
\cite{Monier:etal:2010}.

%%%%%%%%%%%%%%%%%%%%%%%%%%%%%%%%%%%%%%%%%%%%%%%%%%%%%%%%%%%%%%%%%%%%%%%%
\section[NLTE model]{Full NLTE model atmospheres}

Calculation of a full NLTE model atmosphere is a task, which from basic
global stellar parameters like effective temperature (or luminosity),
mass, and radius determines spatial distribution of temperature,
density, ionization states, electron density, velocity, etc.

We would like to emphasize that the resulting model atmospheres may
differ significantly from the LTE solution,
as has already been clearly shown by \cite{Auer:Mihalas:1969:1}.

%=======================================================================
\subsection{Solution of a NLTE model atmosphere}
\label{nltesol}

Solution of a NLTE model atmosphere means determination of distributions
of macroscopic quantities in the stellar atmosphere for given global
parameters describing stellar luminosity $L_\ast$, mass $M_\ast$, and
radius $R_\ast$.
It is achieved via a solution of a system of equations describing
physical properties of the atmosphere.
We demonstrate these equations for a specific case of the static
plane-parallel NLTE model atmosphere.
These equations include the equation of radiative transfer, which
determines the radiation field described using its specific intensity
$I_\mn$,
\begin{equation}
\mu \deriv{I_\mn}{z}
= \eta_\nu - \chi_\nu I_\mn,
\end{equation}
the equations of statistical equilibrium determining atomic level
populations $n_i$,
\begin{equation}
n_i \sum_l \zav{R_{il}+C_{il}} + \sum_l n_l \zav{R_{li}+C_{li}} = 0,
\end{equation}
the equation of hydrostatic equilibrium which determines the density
structure $\rho$,
\begin{equation}
\deriv{p}{m} = g - \frac{4\pi}{c} \int_0^\infty \frac{\chi_\nu}{\rho}
H_\nu \dnu,
\end{equation}
and the equation of radiative equilibrium, which determines the
temperature structure $T$,
\begin{equation}
4\pi\int_0^\infty \zav{\chi_\nu J_\nu - \eta_\nu} \dnu =0.
\end{equation}
In these equations, $n_iR_{il}$ and $n_iC_{il}$ are radiative and
collisional rates for transitions from level $i$ to level $l$,
respectively,
$p$ is the gas pressure,
$g$ is gravitational acceleration,
$c$ is the light speed,
$m$ is the column mass depth ($\dm=-\rho\dz$),
$\rho$ is the
density, $H_\nu$ is the radiative Eddington flux, $J_\nu$ is the mean
radiation intensity,
$\eta_\nu$ and $\chi_\nu$ are emissivity and opacity,
respectively, and $\mu=\cos\theta$ is an angle cosine of a light ray.

This set of equations has to be solved simultaneously.
An efficient method was introduced by \cite{Auer:Mihalas:1969}
as the complete linearization method, which is the multidimensional
Newton-Raphson Method.
If this method is combined with the accelerated lambda iteration method,
significant saving of computing time may be achieved
\citep[e.g.][]{Hubeny:Lanz:1992,Hubeny:Lanz:1995,ATA1,ATA2,ATAsum}.

%=======================================================================
\subsection{Model atmosphere grids}

Since calculation of NLTE model atmospheres is a very time consuming
task even for the relatively simple static \mbox{1-D} case, for practical
purposes it is efficient to use precalculated grids of model atmospheres
and to interpolate between them if a model for required parameters is
not available in the grid.

An example of such a grid is the grid of NLTE model line blanketed
atmospheres of O and B stars \citep{Lanz:Hubeny:2003,Lanz:Hubeny:2007}.
There is also a grid of LTE line blanketed model atmospheres, which
covers almost all reasonable temperatures and gravities, consequently
also hot stars \citep{Kurucz:1993}.

Using a grid instead of a calculation of a model atmosphere we limit
ourselves to several fixed values of grid parameters, which can be for
example the stellar effective temperature {\Teff}, gravitational
acceleration at the stellar surface $g$, stellar radius $R_\star$,
stellar mass $M_\star$, stellar luminosity $L_\star$, and elemental
abundances.
Since the parameter space can be quite extended, using a grid calculated
for a limited range of parameters for model atmosphere analysis may
fail, especially if we want to investigate stars with parameters beyond
the grid ones.
The most efficient possibility is first to use models from a grid to
determine rough values of stellar parameters, which may be then refined
using detailed model atmosphere calculations.
However, if the model grids have sufficiently dense spacing in basic
structural parameters, interpolation of the models may work well, as it
was shown by \cite{Lanz:Hubeny:2003}.

%%%%%%%%%%%%%%%%%%%%%%%%%%%%%%%%%%%%%%%%%%%%%%%%%%%%%%%%%%%%%%%%%%%%%%%%
\section{Comparison of LTE and NLTE modelling}

Since first NLTE model atmospheres were calculated in late 60s
\citep{Auer:Mihalas:1969:1}, there has always been a discussion if NLTE
model atmospheres are really necessary or if LTE models are sufficient.
In any case, the NLTE approximation is more general than the LTE one.
The NLTE effects can be quite complicated.
Consequently, we can only prove that LTE offers acceptable results {\em
after} we calculate a NLTE model, which is able to verify necessary
conditions for LTE.
Assuming LTE means assuming detailed balance in \emph{all} transition a
priori.
Even if we succeed to fit a part of the spectrum with the LTE model
atmosphere, it can not be a proof that the LTE model describes the
atmosphere well.

The basic advantage of LTE models is the fact that they can be
calculated very quickly.
Using contemporary computers we can obtain an LTE model within one
minute or even faster.
On the other hand, calculation of a NLTE model atmosphere may last
several hours or even more.
A question may arise, if we gain anything from these additional hours
of computing time.
Of course, we get a lot.
NLTE model atmosphere calculations give us more accurate level
populations, more accurate ionization balance, more accurate opacities,
more accurate radiation field, and more accurate temperature and density
structure in resulting models.

Although it is clear that the NLTE model atmospheres are superior to the
LTE ones, the latter ones dominated the analysis of hot stars for many
years.
The reason was the problem of line blanketing, where a huge number of
spectral lines in the ultraviolet region caused absorption of
radiation, which was the reemitted in the optical region.
In LTE, relatively straightforward approximations of opacity
distribution function or opacity sampling enabled to handle this effect.
On the other hand, in NLTE model atmospheres the influence of the
radiation field on level populations had to be taken into account, which
was computationally prohibitive until an efficient method enabling
treatment of line blanketing in NLTE was developed.
This method uses the concept of superlevels and superlines
\citep{Anderson:1989}.
Superlevels are averaged atomic energy levels with similar properties,
and superlines are transitions between them.
Details about different applications of this method to calculations of
NLTE model atmospheres can be found in \cite{Dreizler:Werner:1993} and
\cite{Hubeny:Lanz:1995}.
Since now the line blanketing in NLTE can be handled in a satisfactory
manner and since there are also publicly available computer codes able
to solve the problem, e.g. {\TLUSTY}\footnote
{\url{http://nova.astro.umd.edu/index.html}}
\citep{Hubeny:1988,Hubeny:Lanz:1995}
or {\tt PRO2}\footnote{part of
%the T\"ubingen NLTE Model Atmosphere Package
{\tt TMAP}, \url{http://astro.uni-tuebingen.de/~TMAP/}}
\citep{Werner:Dreizler:1999,Werner:etal:2003},
LTE line blanketed model
atmospheres should be replaced by the NLTE ones.

The basic consequence of switching from LTE to NLTE is that
differences in population numbers and also in ionization balance appear.
The LTE populations and ionization fractions are systematically in
error, especially in the outer parts of the atmosphere, which is the
forming region of many spectral lines.
The effect of correct treatment of the equations of kinetic equilibrium
in NLTE model atmospheres is nicely illustrated in Figs. 5-9 of
\cite{Lanz:Hubeny:2003}.
For Rosseland optical depths $\tau_\text{R} \lesssim 1$ the error of the
ionization balance caused by the assumption of LTE is clearly seen.
These errors directly influence the profiles of corresponding lines.
In addition, they also influence heating and cooling in the stellar
atmosphere and lead to differences in the temperature structure, as
illustrated by the Figure 5 in \cite{Lanz:Hubeny:2007}.
At large depths ($\tau_\text{R} \gg 1$), the radiation is close to
isotropic and the diffusion approximation for radiation transfer can be
used, which also means that the radiation is close to its equilibrium
value.
If also the particle velocities are close to the equilibrium
distribution (as is common in ``standard'' stellar atmospheres),
transition rates are very close to the detailed balance
and the microscopic conditions for LTE are fulfilled there.

Since full treatment of NLTE model atmospheres may be computationally
very time consuming, simplified method of solving the NLTE problem
(radiative transfer + rate equations) for selected trace elements
in a \emph{given} model atmosphere is commonly being used.
For more details about this approach we refer the reader to the book by
\cite{Monier:etal:2010}.
Unfortunately, using the LTE model atmospheres instead of the NLTE ones
still dominates this approach.
Besides the availability of the LTE models this is probably caused by
the fact that the temperature structures of LTE and NLTE models are very
similar at large depths.
Some authors advocate using the hybrid method (LTE model atmosphere with
NLTE radiative transfer for trace elements) to be equivalent to the full
NLTE approach.
\cite{Przybilla:etal:2011} even concluded that \emph{`` ... LTE and NLTE
model atmospheres are essentially equivalent for dwarf and giant stars
over the range $15 000\,\Kelvin < \Teff < 35 000\,\Kelvin$, for most
practical applications''}, which can be true if practical applications
are dealing only with lines forming at optical depths
$\tau_\text{R}\gtrsim 1$.
For lines forming above $\tau_\text{R}\approx 1$, which is the majority
of spectral lines, we may expect differences.

Thus, full NLTE modeling (i.e. calculation of NLTE model atmospheres
with the solution of the NLTE problem for trace elements) should be
always preferred since it uses physically consistent assumptions
\citep[see][]{nice2}.
In this approach, the influence of radiation on level occupation
numbers (hence opacities) is not neglected as in the LTE approach.
To be sure that the results are correct, each particular application of
the hybrid LTE/NLTE method has to be independently tested.
We would like to emphasize that in any case, the hybrid LTE/NLTE
approach is significantly superior to pure LTE analysis.
To summarize, using LTE model atmospheres is a fast option, while using
NLTE model atmospheres is a much more exact option, and should be
preferred whenever possible.
The best option to save computing time is to calculate NLTE model
atmosphere as a background model, and then solve the NLTE problem for
trace elements, if the conditions for using trace element approximation
are fulfilled.

%%%%%%%%%%%%%%%%%%%%%%%%%%%%%%%%%%%%%%%%%%%%%%%%%%%%%%%%%%%%%%%%%%%%%%%%
\section{Model atom construction}

An important part of the NLTE calculation is construction of a proper
model of an atom or ion studied.
Some ions are relatively simple, besides hydrogen it is also the
neutral helium \citep[e.g.][]{Auer:Mihalas:1972}.
On the other hand, the total number of levels and corresponding
transitions may be enormous for some ions \citep[see Fig.~5 of][where
the number of transitions is too large to plot,
they form a continuous black surface and one can hardly find any
relevant information from that figure]{Hauschildt:Baron:1995}.

To make the NLTE problem tractable, complicated model atoms may be
simplified.
It is usually done by assuming levels with high quantum number to be in
LTE with respect to the ground level of the next ion.
This way these levels do not enter the equations of statistical
equilibrium.
Another possibility is to merge levels, especially for multiplets
\citep[e.g. for neutral sodium,][]{Gehren:1975,F2rus}.
The most sophisticated way to simplify model atoms is by creating
superlevels, which can be applied to most complicated ions like iron and
nickel.
Superlevels are generalized multiplets, the energy levels of all levels
from a superlevel are averaged (including proper statistical weights),
and similarly the transitions to and from a superlevel members are
averaged.
Then the NLTE line formation problem is solved for model atom consisting
of superlevels.
An example how the superlevels can be created can be found in
\cite{Hubeny:Lanz:1995}.

Collecting data for all transitions is not an easy task for most of
metallic ions.
We have to collect ionization cross sections for all levels included,
transition probabilities for all allowed or forbidden radiative
transitions, and collisional cross sections for all possible
transitions.
If data are not available, we have to look for some reasonable
approximation.
Finally, we have to evaluate values for the merged levels, if they
are used.

%%%%%%%%%%%%%%%%%%%%%%%%%%%%%%%%%%%%%%%%%%%%%%%%%%%%%%%%%%%%%%%%%%%%%%%%
\section{Hot stars with winds}

Stellar wind is an outflow of matter from the stellar surface and it is
a common property of hot massive stars.
The strength of the wind depends on a spectral type, generally, higher
stellar luminosity and lower stellar gravity support stronger stellar
winds.
This means that the strongest winds are for O-type supergiants while for 
dwarf A-type stars the winds are almost absent.

\begin{figure}[t]
\begin{center}
\includegraphics[width=\textwidth]{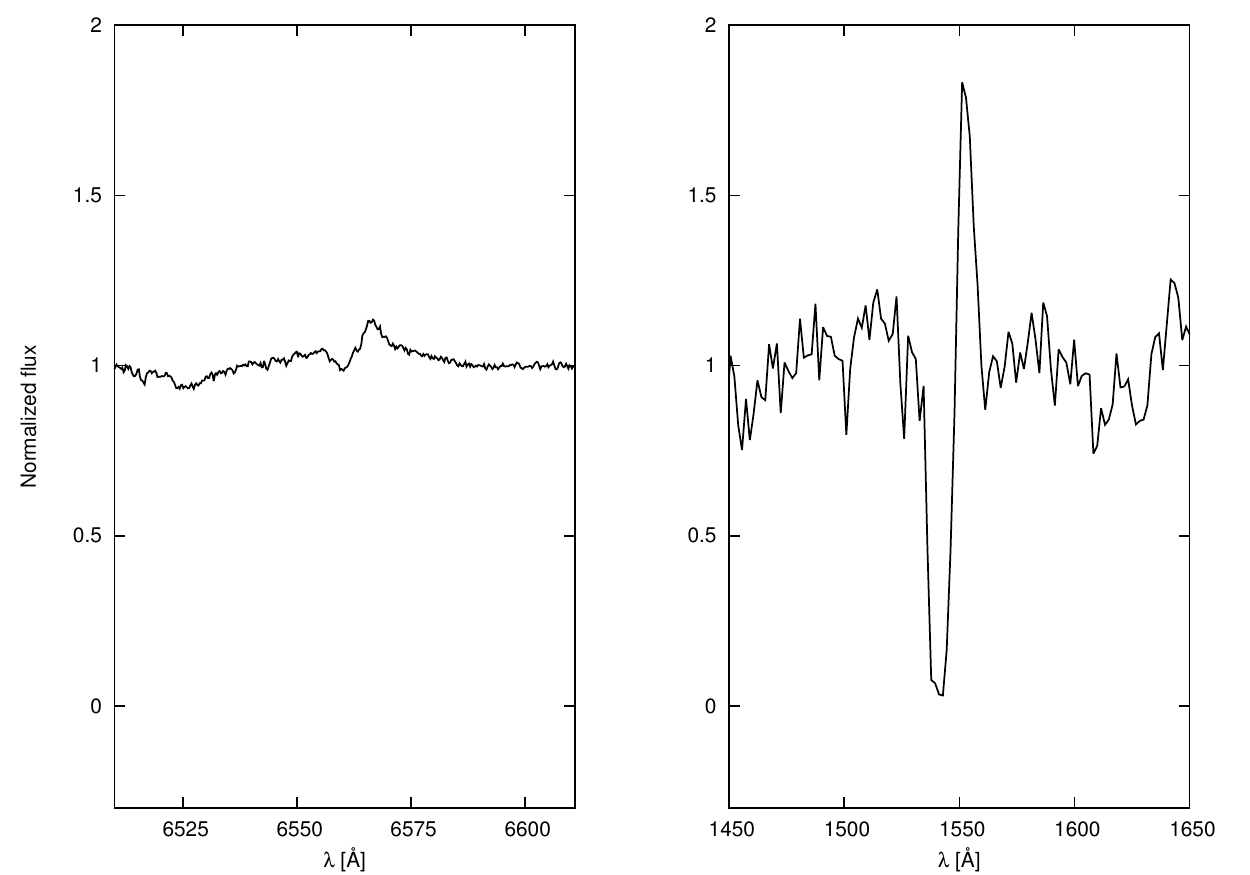}
\end{center}
\caption{Illustration of P-Cygni type profiles (normalised to continuum)
of the line H$\alpha$ (left panel) and the ultraviolet line of
\ion{C}{iv} (right panel) for the star $\lambda$~Cep.
The ultraviolet spectrum is taken from the IUE archive
({\tt http://archive.stsci.edu/iue/}), the visual
spectrum was observed using the Ond\v{r}ejov Perek
Telescope.}\label{pcyg}
\end{figure}

All O-type stars, including O-dwarfs, have stellar wind.
Since these stars emit the maximum of their radiation in the UV
wavelength region, the fact that they lose mass via stellar winds was
discovered by analysis of first ultraviolet spectra of this stellar type
\citep{Morton:1967} obtained by the {\tt Aerobee} rocket.
Observed spectra showed strong P-Cygni type line profiles, for such
lines as of \ion{C}{iv} and \ion{Si}{iv}.
For an example of P-Cygni type line profiles see in Fig.\,\ref{pcyg}.

For hot massive stars, the mass-loss rates ($\dMdtl$), which describe
how quickly the star loses its mass, reach values up to $10^{-6}\,\msy$.
Terminal wind velocities grow from the photosphere and reach their
maximum $v_{\infty}\sim 3000\,\kms$ far from the stellar surface.
The outflow is driven by radiation, which is absorbed or scattered in
continuum (by electrons and by bound-free and free-free transitions) and
in spectral lines.
The dominant contribution to radiative acceleration in these atmospheres
comes from resonance ultraviolet lines of metals, where large flux in
the ultraviolet part of spectrum meets large absorption in these lines,
which is amplified by the Doppler shift caused by the velocity gradient.
On the other hand, the most abundant elements, hydrogen and helium, gain
negligible radiative force.
The momentum gained by metals is transferred to hydrogen and helium by
means of Coulomb collisions between charged ions of metals and
hydrogen-helium component.
Since winds are present practically in all O-type stars and supergiant B
and A-type stars, they have to be taken into account in analysis of hot
stars.

%=======================================================================
\subsection{Modeling of hot stars with winds}

The presence of an outflow is quite a complication in calculations of
consistent models, which can predict theoretical spectra.
To understand how radiation emerging from the stellar photosphere
(static medium) changes passing through their expanding atmosphere, i.e.
wind (moving medium), radiative transfer in both media has to be solved.
There is a number of methods and computer codes for static media
available, but they are not adequate to treat expanding atmospheres of
hot stars.
The problem of the transfer of radiation in moving media is more
complicated, since the wind is expanding, and therefore the Doppler
shift of the photon frequencies must be taken into account.
Absorption and emission coefficients become angle dependent.
Consequently, these coefficients become anisotropic and the aberration
of light may become important \citep[the latter effect is usually
neglected for stellar winds where $\velocity /c \lesssim 0.01$, see][]
{Mihalas:etal:1976}.

There are two basic classes of methods \citep[see, e.g.,][]
{Mihalas:Kunasz:1986} for solving the radiative transfer equation in
moving medium, namely the solution in the observer frame and the
solution in the comoving frame.
In the observer frame, the computations are straightforward, but the
opacity and emissivity become angle dependent.
Consequently, the number of frequency and angle points necessary to
solve the radiative transfer equation may become enormous making the
task computationally expensive.
On the other hand, since the comoving frame is the local rest frame of
the matter, opacity and emissivity are isotropic, but the expressions
for solution of the radiative transfer are more complicated.

To simplify treatment of the problem, a {\em core-halo approximation} is
commonly used, which means that photosphere and wind are modeled
separately, and that the wind does not influence photosphere, while
photospheric flux is a lower boundary condition for wind solution.
Usually, the photosphere is assumed to be static and a full NLTE
line-blanketed model atmosphere modeling described in the
Section~\ref{nltesol} is done.
Here NLTE modeling is inevitable, since LTE is not valid in the
photospheres of O stars.
Then the photospheric flux is taken as a lower boundary condition for
wind solution, which is usually performed for given velocity, density,
and temperature structure, i.e. the NLTE line formation problem,
solution of equations of statistical equilibrium together with the
radiative transfer equation.
Due to the large velocity gradients present in winds of these stars,
Sobolev approximation may be used, and the radiative transfer is
significantly simplified, since then the system of equations of
statistical equilibrium and radiative transfer is local.

With this model, the analysis is done in steps.
The lines which are expected not to be influenced by wind are selected
(the photospheric lines).
These lines serve for $\Teff$ determination, similarly to static models.
Then the wind line profiles are calculated for given velocity
$\velocity(r)$ and density $\rho(r)$ and the mass-loss rate is
determined.
Illustrative examples of application of such a procedure may be found in
\cite{Bouret:etal:2012}.
The most important computer codes for modeling of hot star winds are
listed elsewhere in this book \citep{ja3}. 
A more general background of the analysis of hot stars with winds may be
found, e.g. in \cite{Puls:etal:2008}.

%%%%%%%%%%%%%%%%%%%%%%%%%%%%%%%%%%%%%%%%%%%%%%%%%%%%%%%%%%%%%%%%%%%%%%%%
\section{Summary}

For hot stars NLTE analysis is necessary.
It may be done using either LTE or better NLTE model atmospheres
followed by NLTE for trace elements, if required.
NLTE model atmospheres should be preferred, since they are more exact.
The model atoms have to be carefully constructed.
Model grids may save computing time.
The systematic influence of stellar wind on emergent radiation has to be
taken into account.

\begin{acknowledgement}
J.K. would like to thank Dr. Ewa Niemczura for inviting him to the
Spring School and he would also like to apologize her for the delay in
delivering manuscripts.
B.\v{S}. thanks to Ministry of Education and Science of Republic of
Serbia who supported this work through the project 176002 ``Influence of
collisions on astrophysical plasma spectra''.
The authors are also grateful to all referees (Ivan Hubeny and both
anonymous ones) for their invaluable comments.
This work was partly supported by the project 13-10589S of the Grant
Agency of the Czech Republic (GA \v{C}R).
\end{acknowledgement}

\nocite{wrspec}
\newcommand{\aj}{Astron. J.}
\newcommand{\aap}{Astron. Astrophys.}
\newcommand{\aapr}{Astron. Astrophys. Rev.}
\newcommand{\apj}{Astrophys. J.}
\newcommand{\apjs}{Astrophys. J. Suppl. Ser.}
\newcommand{\araa}{Ann. Rev. Astron. Astrophys.}
\newcommand{\jqsrt}{J. Quant. Spectrosc. Radiat. Transfer}
\newcommand{\mnras}{Mon. Not. Roy. Astron. Soc.}

\bibliographystyle{spbasic}
%only spbasic works with natbib
%\bibliographystyle{spmpsci}
%\bibliographystyle{spphys}
%\bibliographystyle{astron}
\bibliography{kubat,wrspec,proc}

\begin{thebibliography}{32}
\providecommand{\natexlab}[1]{#1}
\providecommand{\url}[1]{{#1}}
\providecommand{\urlprefix}{URL }
\expandafter\ifx\csname urlstyle\endcsname\relax
  \providecommand{\doi}[1]{DOI~\discretionary{}{}{}#1}\else
  \providecommand{\doi}{DOI~\discretionary{}{}{}\begingroup
  \urlstyle{rm}\Url}\fi
\providecommand{\eprint}[2][]{\url{#2}}

\bibitem[{Anderson(1989)}]{Anderson:1989}
Anderson LS (1989) Line blanketing without local thermodynamic equilibrium.
  {II} - {A} solar-type model in radiative equilibrium. \apj 339:558--578,
  \doi{10.1086/167317}

\bibitem[{Auer and Mihalas(1969{\natexlab{a}})}]{Auer:Mihalas:1969:1}
Auer LH, Mihalas D (1969{\natexlab{a}}) Non-{LTE} model atmospheres. {I}.
  {R}adiative equilibrium models with {L}yman alpha. \apj 156:157,
  \doi{10.1086/149955}

\bibitem[{Auer and Mihalas(1969{\natexlab{b}})}]{Auer:Mihalas:1969}
Auer LH, Mihalas D (1969{\natexlab{b}}) Non-{LTE} model atmospheres. {III}. {A}
  complete-linearization method. \apj 158:641--655, \doi{10.1086/150226}

\bibitem[{Auer and Mihalas(1972)}]{Auer:Mihalas:1972}
Auer LH, Mihalas D (1972) Non-{LTE} model atmospheres. {VII}. {T}he hydrogen
  and helium spectra of the o stars. \apjs 24:193, \doi{10.1086/190253}

\bibitem[{Bouret et~al(2012)Bouret, Hillier, Lanz, and
  Fullerton}]{Bouret:etal:2012}
Bouret JC, Hillier DJ, Lanz T, Fullerton AW (2012) Properties of galactic
  early-type {O}-supergiants. {A} combined {FUV-UV} and optical analysis. \aap
  544:A67, \doi{10.1051/0004-6361/201118594}

\bibitem[{Boyarchuk et~al(1988)Boyarchuk, Hubeny, Kub{\'{a}}t, Lyubimkov, and
  Sakhibullin}]{F2rus}
Boyarchuk AA, Hubeny I, Kub{\'{a}}t J, Lyubimkov LS, Sakhibullin NA (1988)
  Non-{LTE} effects in the atmospheres of {F}-type supergiants - {P}art {T}wo -
  {A}nalysis of {N}a{I} lines (the method of computations). Astrofizika
  28:335--342

\bibitem[{Dreizler and Werner(1993)}]{Dreizler:Werner:1993}
Dreizler S, Werner K (1993) Line blanketing by iron group elements in
  {N}on-{LTE} model atmospheres for hot stars. \aap 278:199--208

\bibitem[{Gehren(1975)}]{Gehren:1975}
Gehren T (1975) Kinetic equilibrium and line formation of {N}a {I} in the solar
  atmosphere. \aap 38:289--302

\bibitem[{Hauschildt and Baron(1995)}]{Hauschildt:Baron:1995}
Hauschildt PH, Baron E (1995) Non-{LTE} treatment of {F}e {II} in astrophysical
  plasmas. \jqsrt 54:987--999, \doi{10.1016/0022-4073(95)00118-5}

\bibitem[{Hubeny(1988)}]{Hubeny:1988}
Hubeny I (1988) A computer program for calculating non-{LTE} model stellar
  atmospheres. Computer Physics Communications 52:103--132,
  \doi{10.1016/0010-4655(88)90177-4}

\bibitem[{Hubeny and Lanz(1992)}]{Hubeny:Lanz:1992}
Hubeny I, Lanz T (1992) Accelerated complete-linearization method for
  calculating {NLTE} model stellar atmospheres. \aap 262:501--514

\bibitem[{Hubeny and Lanz(1995)}]{Hubeny:Lanz:1995}
Hubeny I, Lanz T (1995) Non-{LTE} line-blanketed model atmospheres of hot
  stars. 1: Hybrid complete linearization/accelerated lambda iteration method.
  \apj 439:875--904, \doi{10.1086/175226}

\bibitem[{Hubeny and Mihalas(2014)}]{Hubeny:Mihalas:2014}
Hubeny I, Mihalas D (2014) Theory of Stellar Atmospheres. Princeton University
  Press, Princeton and Oxford, in press

\bibitem[{Kub{\'{a}}t(1994)}]{ATA1}
Kub{\'{a}}t J (1994) Spherically symmetric model atmospheres using approximate
  lambda operators {I}: {F}irst results for static {NLTE} atmospheres. {\aap}
  287:179--190

\bibitem[{Kub{\'{a}}t(1996)}]{ATA2}
Kub{\'{a}}t J (1996) Spherically symmetric model atmospheres using approximate
  lambda operators {II}. {S}imple method for calculation of both plane-parallel
  and spherically symmetric static model atmospheres. {\aap} 305:255--264

\bibitem[{Kub{\'{a}}t(2003)}]{ATAsum}
Kub{\'{a}}t J (2003) Calculation of spherically symmetric {NLTE} model
  atmospheres using {ALI} and a thermal balance method. In: Piskunov N, Weiss
  WW, Gray DF (eds) Modelling of Stellar Atmospheres, IAU Symposium, vol 210,
  p~A8

\bibitem[{Kub{\'{a}}t(2010)}]{nice2}
Kub{\'{a}}t J (2010) Statistical equilibrium equations for trace elements in
  stellar atmospheres. In: Monier R, Smalley B, Wahlgren G, Stee P (eds)
  Non-LTE Line Formation for Trace Elements in Stellar Atmospheres, EDP
  Sciences, EAS Publications Series, vol~43, pp 43--54,
  \doi{10.1051/eas/1043004}

\bibitem[{Kub{\'{a}}t(2014)}]{ja3}
Kub{\'{a}}t J (2014) Current status of {NLTE} analysis. In:  \cite{wrspec},
  these proceedings

\bibitem[{Kurucz(1993)}]{Kurucz:1993}
Kurucz R (1993) {ATLAS9} stellar atmosphere programs and 2 km/s grid. In:
  Kurucz CD-ROM, vol~13, Smithsonian Astrophysical Observatory, Cambridge,
  Mass.

\bibitem[{Lanz and Hubeny(2003)}]{Lanz:Hubeny:2003}
Lanz T, Hubeny I (2003) A grid of non-{LTE} line-blanketed model atmospheres of
  {O}-type stars. \apjs 146:417--441, \doi{10.1086/374373}

\bibitem[{Lanz and Hubeny(2007)}]{Lanz:Hubeny:2007}
Lanz T, Hubeny I (2007) A grid of {NLTE} line-blanketed model atmospheres of
  early {B}-type stars. \apjs 169:83--104, \doi{10.1086/511270}

\bibitem[{Lanz et~al(2008)Lanz, Cunha, Holtzman, and Hubeny}]{Lanz:etal:2008}
Lanz T, Cunha K, Holtzman J, Hubeny I (2008) Argon abundances in the solar
  neighborhood: Non-{LTE} analysis of orion association {B}-type stars. \apj
  678:1342--1350, \doi{10.1086/587047}

\bibitem[{Mihalas(1978)}]{Mihalas:1978}
Mihalas D (1978) Stellar atmospheres, 2nd edn. W. H. Freeman \& Co., San
  Francisco

\bibitem[{Mihalas and Kunasz(1986)}]{Mihalas:Kunasz:1986}
Mihalas D, Kunasz PB (1986) The computation of radiation transport using
  {F}eautrier variables. {II} - {S}pectrum line formation in moving media.
  Journal of Computational Physics 64:1--26, \doi{10.1016/0021-9991(86)90016-1}

\bibitem[{Mihalas et~al(1976)Mihalas, Kunasz, and Hummer}]{Mihalas:etal:1976}
Mihalas D, Kunasz PB, Hummer DG (1976) Solution of the comoving-frame equation
  of transfer in spherically symmetric flows {III}. {E}ffect of aberration and
  advection terms. \apj 206:515--524, \doi{10.1086/154407}

\bibitem[{Monier et~al(2010)Monier, Smalley, Wahlgren, and
  Stee}]{Monier:etal:2010}
Monier R, Smalley B, Wahlgren G, Stee P (eds) (2010) Non-LTE Line Formation for
  Trace Elements in Stellar Atmospheres, EAS Publications Series, vol~43

\bibitem[{Morton(1967)}]{Morton:1967}
Morton DC (1967) The far-ultraviolet spectra of six stars in {O}rion. \apj
  147:1017, \doi{10.1086/149091}

\bibitem[{Niemczura et~al(2014)Niemczura, Smalley, and Pych}]{wrspec}
Niemczura E, Smalley B, Pych W (eds) (2014) Spring School of Spectroscopic Data
  Analyses, GeoPlanet: Earth and Planetary Sciences, Springer Verlag, Berlin,
  these proceedings

\bibitem[{Przybilla et~al(2011)Przybilla, Nieva, and
  Butler}]{Przybilla:etal:2011}
Przybilla N, Nieva MF, Butler K (2011) Testing common classical {LTE} and
  {NLTE} model atmosphere and line-formation codes for quantitative
  spectroscopy of early-type stars. Journal of Physics Conference Series
  328(1):012015, \doi{10.1088/1742-6596/328/1/012015}

\bibitem[{Puls et~al(2008)Puls, Vink, and Najarro}]{Puls:etal:2008}
Puls J, Vink JS, Najarro F (2008) Mass loss from hot massive stars. \aapr
  16:209--325, \doi{10.1007/s00159-008-0015-8}

\bibitem[{Werner and Dreizler(1999)}]{Werner:Dreizler:1999}
Werner K, Dreizler S (1999) The classical stellar atmosphere problem. Journal
  of Computational and Applied Mathematics 109:65--93

\bibitem[{Werner et~al(2003)Werner, Deetjen, Dreizler, Nagel, Rauch, and
  Schuh}]{Werner:etal:2003}
Werner K, Deetjen JL, Dreizler S, Nagel T, Rauch T, Schuh SL (2003) Model
  photospheres with accelerated lambda iteration. In: Hubeny I, Mihalas D,
  Werner K (eds) Stellar Atmosphere Modeling, Astronomical Society of the
  Pacific Conference Series, vol 288, p~31

\end{thebibliography}

\end{document}